\documentclass[superscriptaddress,groupedaddress,nofootnoteinbib,12pt]{article}  % for review and submission

\usepackage{amsmath}
\usepackage{amsfonts}
\usepackage{authblk}
\usepackage{bbm}
\usepackage{color}
\usepackage[dvipsnames]{xcolor}
\usepackage{graphicx,graphics}
\usepackage{epstopdf}
\usepackage{caption}
\usepackage{subcaption}
\usepackage{float} %This package is needed for the positioning the graphicx
\usepackage{pifont}
\usepackage{titlesec}
\usepackage{etoolbox}
\usepackage{indentfirst}
\usepackage{jcappub}
\usepackage{epstopdf}

\def\be{\begin{equation}}
\def\ee{\end{equation}}
\def\ba{\begin{array}}
\def\ea{\end{array}}
\def\babc{\begin{subequations}}
\def\eabc{\end{subequations}}

\def\5{\hspace*{5mm}}
\def\2{{\scriptstyle\frac12}}

\begin{document}

\title{Possible Derivative Interactions in Massive bi-Gravity}

\author{Giorgi Tukhashvili}
\affiliation{Center for Cosmology and Particle Physics, Department of Physics, \\
New York University, New York, NY, 10003, USA}

\emailAdd{gt783@nyu.edu}


\abstract{Using the metric formalism, we study the derivative mixings of spin-2 fields in massive bi-Gravity. Necessary (but not sufficient) criteria are given for such mixings to be ghost free. Examples satisfying those criteria are studied and it is shown that in the decoupling limit they host a ghost.}

\maketitle

%%%%%%%%%%%%%%%%%%%%%%%%%%%%%%%%%%%%%%%%%%%%%%%%%%%%%%%%%%%%%%%%%%%%%%%%%%
%%%%%%%%%%%%%%%%%%%%%%%%%%%%%%%%%%%%%%%%%%%%%%%%%%%%%%%%%%%%%%%%%%%%%%%%%%
%%%%%%%%%%%%%%%%%%%%%%%%%%%%%%%%%%%%%%%%%%%%%%%%%%%%%%%%%%%%%%%%%%%%%%%%%%

\section{Introduction}

%%%%%%%%%%%%%%%%%%%%%%%%%%%%%%%%%%%%%%%%%%%%%%%%%%%%%%%%%%%%%%%%%%%%%%%%%%
%%%%%%%%%%%%%%%%%%%%%%%%%%%%%%%%%%%%%%%%%%%%%%%%%%%%%%%%%%%%%%%%%%%%%%%%%%
%%%%%%%%%%%%%%%%%%%%%%%%%%%%%%%%%%%%%%%%%%%%%%%%%%%%%%%%%%%%%%%%%%%%%%%%%%

In 1940, N. Rosen published a pair of papers \cite{Rosen:1940zza,Rosen:1940zz} in where he proposed a view on gravitation different from Einstein's General Theory of Relativity. Per Rosen's interpretation, the tensor, $g_{\mu \nu}$,
which describes the gravitational field, has no connection with geometry and it propagates on a flat background described by the tensor $\tilde{g}^0_{\mu \nu}$. By flat we mean that the corresponding Riemann tensor vanishes,
$R^{\lambda}_{\mu \nu \rho} (\tilde{g}^0)=0$\footnote{The most general parametrization of such a metric can be done as $\tilde{g}^0_{\mu \nu} = \partial_\mu \Phi^a \partial_\nu \Phi^b \eta_{a b}$. Where $\eta_{a b}$ is the Minkowski metric.}. One should note that both
objects, $ g_{\mu \nu}$ and $\tilde{g}^0_{\mu \nu}$, are tensors with respect to the same General Coordinate Transformations ($diff$). It turns out that such a formalism has some advantages and the predictions, in general,
coincide with Einstein's General Theory of Relativity (GR). This model is defined by the following action:
\be
S = \int d^4 x  \sqrt{g}  g^{\mu \nu}\left( \Delta^\rho_{\nu \lambda} \Delta^\lambda_{\mu \rho} - \Delta^\rho_{\lambda \rho} \Delta^\lambda_{\mu \nu} \right).
\ee
\be
\Delta^\lambda_{\mu \nu} \equiv \Gamma^\lambda_{\mu \nu} (g) - \Gamma^\lambda_{\mu \nu} (\tilde{g}^0).
\ee
This was the very first attempt to create a bi-metric theory of gravity. Flatness of
$\tilde{g}^0_{\mu \nu}$ is a crucial part of this theory, since promoting $\tilde{g}^0_{\mu \nu}$ to a dynamical field would generate a ghost even at the linear level. \\
\indent The study of the interacting spin-2 fields dates back to 1958, when H. A. Buchdahl \cite{Buchdahl:1958xv} studied the interaction between the gravity and higher spin $(>3/2)$ fields.
He argued that these interactions were strongly constrained and precluded any interesting solutions. C. Aragone and S. Deser \cite{Aragone:1971kh} showed that the inconsistency was caused by the broken $diff$ invariance.
The result was generalized by N. Boulanger \textit{et. al.} \cite{Boulanger:2000rq} who showed that, ``in the massless case, there is no ghost free coupling, with at most two derivatives of the fields,
that can mix different spin-2 fields''. In other words, the most general action for massless spin-2 fields is a sum of Einstein-Hilbert terms:
\be
S = \sum_{n} \frac{1}{2 M^2_{n}}\int d^4 x \sqrt{-g^{[n]}} R^{[n]}.
\ee
Interactions through the mass term were introduced by C. J. Isham, A. Salam, and J. Strathdee (ISS) \cite{Isham:1971gm}, who developed the first theory of massive bi-Gravity by analogy with the Vector
Meson Dominance model \cite{Kroll:1967it}. By then the correct form of the mass term was not known, so the ISS model hosts the Boulware-Deser (BD) ghost \cite{Boulware:1973my}. After the discovery of the ghost free massive
gravity \cite{deRham:2010ik,deRham:2010kj}, the ISS model was reformulated in \cite{Hassan:2011zd} and it was shown to be ghost free. \\
\indent As we already mentioned, in the massless case, the presence of the ghost in the mixing terms is due to the broken diffeomorphism invariance. For the massive case, in order to make this invariance explicit, one needs to introduce the St\"uckelberg fields in the de Rham-Gabadadze-Tolley (dRGT) \cite{deRham:2010ik,deRham:2010kj} potential. We think that these new degrees of freedom might help to avoid the ghost coming from the derivative interactions ($\equiv$ kinetic mixings) as well. There were no successful attempts to construct such terms, both in massive gravity and massive bi-Gravity \cite{Kimura:2013ika,deRham:2013tfa,deRham:2015rxa}.\footnote{See \cite{Hinterbichler:2013eza} for a potential loophole.} \\

Within this paper, we will work out the necessary (but not sufficient) criteria for the kinetic mixings to be ghost free. Then we will build terms that satisfy those criteria and we'll try to study them using the following examples:
\be\label{addition}
\frac{M_h M_f}{2} \sqrt{-g} g^{\mu \nu} \tilde{R}_{\mu \nu}  +
   \beta  \frac{M_h M_f}{2} \sqrt{-\tilde{g}} \tilde{g}^{\mu \nu} R_{\mu \nu}.
\ee
We will show that in the $\Lambda = \left( m^2 M_p \right)^{1/3}$ decoupling limit, (\ref{addition}) generates both strongly coupled terms and a ghost. \\
\indent In general, conventions through the paper coincide with those of \cite{deRham:2014zqa}. To denote the strong coupling scale, we will use $\Lambda = \left( m^2 M_p \right)^{1/3}$ instead of usual convention $\Lambda_3$.
The objects (Christoffel symbol, Ricci tensor, etc.) with tilde are defined with respect to (wrt) the metric $\tilde{g}_{\mu \nu}$, while those without tilde wrt $g_{\mu \nu}$. Minkowski metric is mostly minus.

%%%%%%%%%%%%%%%%%%%%%%%%%%%%%%%%%%%%%%%%%%%%%%%%%%%%%%%%%%%%%%%%%%%%%%%%%%
%%%%%%%%%%%%%%%%%%%%%%%%%%%%%%%%%%%%%%%%%%%%%%%%%%%%%%%%%%%%%%%%%%%%%%%%%%
%%%%%%%%%%%%%%%%%%%%%%%%%%%%%%%%%%%%%%%%%%%%%%%%%%%%%%%%%%%%%%%%%%%%%%%%%%

\section{Extended Bi-Gravity}

%%%%%%%%%%%%%%%%%%%%%%%%%%%%%%%%%%%%%%%%%%%%%%%%%%%%%%%%%%%%%%%%%%%%%%%%%%
%%%%%%%%%%%%%%%%%%%%%%%%%%%%%%%%%%%%%%%%%%%%%%%%%%%%%%%%%%%%%%%%%%%%%%%%%%
%%%%%%%%%%%%%%%%%%%%%%%%%%%%%%%%%%%%%%%%%%%%%%%%%%%%%%%%%%%%%%%%%%%%%%%%%%

\begin{figure}[t]
    \centering
    \begin{subfigure}[b]{0.4\textwidth}
        \includegraphics[width=\textwidth]{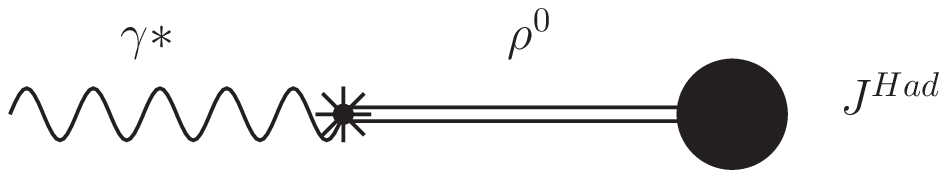}
        \caption{Sketch of the Vector Meson Dominance model.}
        \label{FIG_VMD}
    \end{subfigure}
    \begin{subfigure}[b]{0.4\textwidth}
        \includegraphics[width=\textwidth]{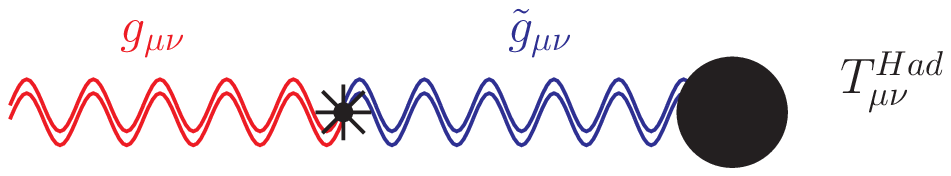}
        \caption{Sketch of the $f$ Dominance model.}
        \label{FIG_TMD}
    \end{subfigure}
    \caption{(a) Off-shell photon decaying into the $\rho^0$ meson which couples to the hadronic current. By integrating out the $\rho^0$ meson one can generate the electromagnetic form-factor for a given hadron. (b) Graviton (red)
decaying into massive spin-2 particle (blue) which couples to the hadronic energy-momentum tensor.}
\end{figure}

The first theory of massive Bi-Gravity, the $f$ Dominance or Tensor Meson Dominance was introduced by C. J. Isham, A. Salam, and J. Strathdee \cite{Isham:1971gm}. The idea is based on the Vector Meson Dominance (VMD)
model \cite{Kroll:1967it,VMD}. Within the VMD model the electromagnetic field couples directly to leptons and through a vector meson to hadrons (Fig. \ref{FIG_VMD}).
The VMD model successfully managed to qualitatively explain the hadronic form factors.
In a complete analogy with VMD, \textit{C. J. Isham et. al.} \cite{Isham:1971gm} postulated the existence of a new spin-2 massive particle, $\tilde{g}_{\mu \nu}$, that was coupled to the hadrons, while the graviton, $g_{\mu \nu}$, was coupled to
leptons (Fig. \ref{FIG_TMD}). On the Lagrangian level this theory can be expressed as:
\begin{align}
\nonumber  \mathcal{L}_{ISS} = & - \frac{1}{2 M^2_h} \sqrt{-g} g^{\mu \nu} R_{\mu \nu} + \mathcal{L} \left( g_{\mu \nu} , \text{leptons} \right) - \\
  {} & - \frac{1}{2 M^2_f} \sqrt{-\tilde{g}} \tilde{g}^{\mu \nu} \tilde{R}_{\mu \nu} + \mathcal{L} \left( \tilde{g}_{\mu \nu} , \text{hadrons} \right)
  + \mathcal{L}_{mass} \left( g_{\mu\nu} , \tilde{g}_{\mu\nu} \right).
\end{align}
The mixing of the fields happens through the mass term, $\mathcal{L}_{mass}$, without which the two worlds do not communicate with each other. Choosing this term properly is a crucial step in the model. The original choice
for $\mathcal{L}_{mass}$ by \textit{C. J. Isham et. al.} exhibits the BD ghost. Recently a ghost free form of $\mathcal{L}_{mass}$ was found by C. de Rham, G. Gabadadze and A. J. Tolley \cite{deRham:2010ik,deRham:2010kj} and
this opened new frontiers for massive bi-Gravity. Using the new mass term, to which we will refer as $\mathcal{L}_{dRGT}$, the ISS model was reformulated in \cite{Hassan:2011zd} and it was shown to be ghost
free. The spectrum of this model consists of one massive and one massless spin-2 fields i.e. 7 healthy degrees of freedom (DoF). After including the dRGT mass term we need new DoF to recover the full diffeomorphism invariance, $diff \otimes \widetilde{diff}$, one might expect that these new DoF could help to avoid the ghost coming from the kinetic mixings as well. The following criteria need to be satisfied for the kinetic mixings to be ghost free ($h$ and $f$ are the fluctuations of $g$ and $\tilde{g}$ around Minkowski background respectively):
 \begin{enumerate}
   \item On the linear level generate $(\partial h)^2$, $(\partial f)^2$ and $\partial h \partial f$. \label{crit1}
   \item Have at least one copy of $diff$ invariance nonlinearly. \label{crit2}
   \item On the linear level have two copies of $diff$ invariance. \label{crit3}
 \end{enumerate}
\textbf{These criteria are necessary but not sufficient for the ghost to be absent.} Let's discuss them in more detail. Criterion \ref{crit1} is obvious since we are building the kinetic mixings. \ref{crit2} assures that we get a scalar under the diagonal subgroup, $diag(diff \otimes \widetilde{diff})$, and we can use the same St\"uckelberg fields to recover the two copies of diffeomorphism invariance, both, in the mass term and in the kinetic mixings. In other words: \ref{crit2} helps us to avoid new DoF. Criterion \ref{crit3} guarantees that, up to a field redefinition, on the linear level we get Einstein-Hilbert action, which is unique to avoid Ostrogradski instability.

It is easy to find the building blocks of this model. The object that transforms covariantly under $diff$ and generates terms like $(\partial h)^2$, $(\partial f)^2$ on the linear level is the Riemann tensor. In addition to the Riemann tensor we have $\Delta^\rho_{\mu \nu} = \Gamma^\rho_{\mu \nu} - \tilde{\Gamma}^\rho_{\mu \nu}$ which transforms covariantly under $diag(diff \otimes \widetilde{diff})$, so terms like $\partial \Delta$ and $\Delta \Delta$ are also allowed.\footnote{Following relation holds: $\sqrt{-g} g^{\mu \nu} R_{\mu \nu}= \partial_\mu \left( \right)^\mu + \sqrt{-g} g^{\mu \nu} \tilde{R}_{\mu \nu} + \sqrt{-g} g^{\mu \nu} \left( \Delta^\rho_{\nu \lambda} \Delta^\lambda_{\mu \rho} - \Delta^\rho_{\lambda \rho} \Delta^\lambda_{\mu \nu}\right)$. This is a general version of the formula given in \cite{Rosen:1940zza}.} All we need is to build scalars out of these tensors. There exist infinitely many terms that satisfy these criteria. Here are some of them:
\be\label{k1}
\sqrt{-g} g^{\mu \nu} \tilde{R}_{\mu \nu},  \5\5\5  \sqrt{-\tilde{g}} \tilde{g}^{\mu \nu} R_{\mu \nu},
\ee
\be\label{k2}
\sqrt{-\tilde{g}} g^{\mu \nu} R_{\mu \nu} + \sqrt{-g} \tilde{g}^{\mu \nu} R_{\mu \nu},
\ee
\be\label{k3}
\sqrt{-\tilde{g}} \tilde{g}^{\alpha \mu} \tilde{g}^{\beta \nu} R_{\alpha \beta \mu \nu} - \sqrt{-g} \tilde{g}^{\alpha \mu} \tilde{g}^{\beta \nu} R_{\alpha \beta \mu \nu}
 + \sqrt{-g} \tilde{g}^{\mu \nu} R_{\mu \nu},
\ee
\be\label{k4}
\sqrt{-g} \tilde{g}^{\mu \alpha_1} g_{\alpha_1 \beta_1} \tilde{g}^{\beta_1 \alpha_2} \cdots g_{\alpha_n \beta_n} \tilde{g}^{\beta_n \nu} R_{\mu \nu} - (n+1) \sqrt{-g} \tilde{g}^{\mu \nu} R_{\mu \nu},
\ee
\be
\nonumber \cdots
\ee
The particular combinations in (\ref{k2}),(\ref{k3}),(\ref{k4}) are chosen in order to satisfy criterion \ref{crit3}.\footnote{On the linear level $\sqrt{-g} g^{\mu \nu} \tilde{R}_{\mu \nu}= - h^{\mu \nu} \tilde{G}^{(1)}_{\mu \nu} + \eta^{\mu \nu} \tilde{R}^{(2)}_{\mu \nu}$, which obviously has two copies of $diff$ invariance due to the Bianchi identity ($\tilde{G}^{(1)}_{\mu \nu}$ and $\tilde{R}^{(2)}_{\mu \nu}$ are the linearized Einstein tensor and quadratic Ricci tensor respectively). Same is true for the other terms.} Because these terms are healthy at the linear level and it's possible to recover the nonlinear $diff$ invariance without introducing new DoF, we think that it is worthy to study their nonlinear effects as well. \\
\indent We will try to study the new terms by considering the example of (\ref{k1}). From now on, our starting point will be the following action:
\begin{align}\label{EbG}
\nonumber  S_{EbG} = \int d^4 x  & \left[ \frac{M_h^2}{2} \sqrt{-g} g^{\mu \nu} R_{\mu \nu} +
(\alpha - \beta ) \frac{M_h M_f}{2} \sqrt{-g} g^{\mu \nu} \tilde{R}_{\mu \nu} + \right. \\
  {} &  + \beta  \frac{M_h M_f}{2} \sqrt{-\tilde{g}} \tilde{g}^{\mu \nu} R_{\mu \nu} +
  \gamma \frac{M_f^2}{2} \sqrt{-\tilde{g}} \tilde{g}^{\mu \nu} \tilde{R}_{\mu \nu} + \\
\nonumber  {} &  \left. + \frac{m^2 M_{h}^2}{4} \sqrt{-g} \left( \mathcal{L}_2 [\mathcal{K}] +
\alpha_3 \mathcal{L}_3 [\mathcal{K}] + \alpha_4 \mathcal{L}_4 [\mathcal{K}] \right) + \mathcal{L} \left( g_{\mu \nu}, \text{matter fields} \right) \right],
\end{align}
where the term in the parenthesis in the third line is the usual dRGT mass term with $\mathcal{K}^\alpha_\beta = \delta^\alpha_\beta - \sqrt{g^{\alpha \lambda} \tilde{g}_{\lambda \beta}}$ and for simplicity only one metric is coupled to matter. We will refer to the model defined by (\ref{EbG}) as \emph{Extended Bi-Gravity (EbG)}
and assume that $ \alpha \neq \beta \neq 0$. To the best of our knowledge this model has not been studied before.\footnote{In terms of Tetrads $\int d^4 x \sqrt{-\tilde{g}} \tilde{g}^{\mu \nu} R_{\mu \nu} = \frac{1}{2} \int \varepsilon_{\mathbbm{a b c d}} f_\alpha^\mathbbm{a} f^{\beta \mathbbm{b}} e_a^\alpha e_{\beta b} R^{a b} \wedge f^\mathbbm{c} \wedge f^\mathbbm{d}$ and $\int d^4 x \sqrt{-g} g^{\mu \nu} \tilde{R}_{\mu \nu} = \frac12 \int \varepsilon_{a b c d} e_\alpha^a e^{\beta b} f_\mathbbm{a}^\alpha f_{\beta \mathbbm{b}} \tilde{R}^{\mathbbm{a b}} \wedge e^c \wedge e^d$ ($e$ and $f$ are tetrads corresponding to metrics $g$ and $\tilde{g}$ respectively) which are different from those discussed in \cite{deRham:2015rxa}: $\int \varepsilon_{a b c d} R^{a b} \wedge \left( e^c \wedge f^d + \beta f^c \wedge f^d \right) $. Although they break the $diff \otimes \widetilde{diff}$ to the diagonal, the new terms preserve the two copies of Local Lorentz Transformations. }

%%%%%%%%%%%%%%%%%%%%%%%%%%%%%%%%%%%%%%%%%%%%%%%%%%%%%%%%%%%%%%%%%%%%%%%%%%
%%%%%%%%%%%%%%%%%%%%%%%%%%%%%%%%%%%%%%%%%%%%%%%%%%%%%%%%%%%%%%%%%%%%%%%%%%
%%%%%%%%%%%%%%%%%%%%%%%%%%%%%%%%%%%%%%%%%%%%%%%%%%%%%%%%%%%%%%%%%%%%%%%%%%

\subsection{The Linear Theory}

%%%%%%%%%%%%%%%%%%%%%%%%%%%%%%%%%%%%%%%%%%%%%%%%%%%%%%%%%%%%%%%%%%%%%%%%%%
%%%%%%%%%%%%%%%%%%%%%%%%%%%%%%%%%%%%%%%%%%%%%%%%%%%%%%%%%%%%%%%%%%%%%%%%%%
%%%%%%%%%%%%%%%%%%%%%%%%%%%%%%%%%%%%%%%%%%%%%%%%%%%%%%%%%%%%%%%%%%%%%%%%%%

\begin{figure}[H]
\centering
\includegraphics[width=0.7\textwidth]{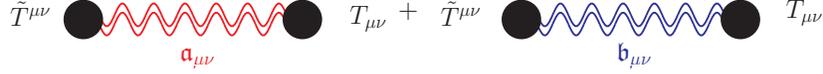}
\caption{Feynman diagram for the exchange amplitude between two sources. Red: $\mathfrak{a}_{\mu \nu}$ contribution, Blue: $\mathfrak{b}_{\mu \nu}$ contribution.}\label{feyn}
\end{figure}

Now we'll investigate the linear limit of (\ref{EbG}) and we'll show that in this limit there are one massive and one massless spin-2 modes, i.e. seven healthy degrees of freedom. \\
\indent Let's expand the fields around the Minkowski background, with $h_{\mu \nu}$ and $f_{\mu \nu}$ being the fluctuations of $g_{\mu \nu}$ and $\tilde{g}_{\mu \nu}$ respectively. Up to quadratic order in fields (\ref{EbG}) takes the form:
\begin{align}\label{LinEbG}
\nonumber  \mathcal{S}_{EbG} = \int d^4 x & \left[ - \frac{1}{4}
\left( \begin{array}{c} h^{\mu \nu} \\ f^{\mu \nu} \end{array} \right)^T
\left( \begin{array}{cc} 1 - \beta/ \kappa & \alpha \\
\alpha & \gamma - \alpha \kappa + \beta \kappa \end{array} \right) \mathcal{E}^{\alpha \beta}_{\mu \nu}
\left( \begin{array}{c} h_{\alpha \beta} \\ f_{\alpha \beta} \end{array} \right) - \right. \\
  {} &  \left. - \frac{m^2}{4} \left( \begin{array}{c} h^{\mu \nu} \\ f^{\mu \nu} \end{array} \right)^T
\left( \begin{array}{cc} 1  & -\kappa \\
-\kappa &  \kappa^2 \end{array} \right) \mathcal{F}^{\alpha \beta}_{\mu \nu}
\left( \begin{array}{c} h_{\alpha \beta} \\ f_{\alpha \beta} \end{array} \right) + \frac{1}{2 M_h} h_{\mu \nu} T^{\mu \nu} \right].
\end{align}
\be
\mathcal{E}^{\alpha \beta}_{\mu \nu} \equiv - \frac12 \left(
\delta^\alpha_\mu \delta^\beta_\nu \partial^2 + \eta^{\alpha \beta} \partial_\mu \partial_\nu + \eta_{\mu \nu} \partial^\alpha \partial^\beta
- \eta_{\mu \nu}  \eta^{\alpha \beta} \partial^2 - \delta^\alpha_\nu \partial^\beta \partial_\mu - \delta^\beta_\mu \partial^\alpha \partial_\nu \right),
\ee
\be
\mathcal{F}^{\alpha \beta}_{\mu \nu} \equiv \frac12 \left( \delta_\mu^\alpha \delta_\nu^\beta - \eta_{\mu \nu} \eta^{\alpha \beta} \right).
\ee
In order for both kinetic terms to have the correct sign, we need to constrain the parameters ($\kappa \equiv \frac{M_h}{M_f}>0$):
\be\label{cons}
\kappa > \beta, \5\5 \gamma > \kappa \left( \alpha - \beta + \frac{\alpha^2}{\kappa - \beta} \right).
\ee
After diagonalizing kinetic and mass terms and canonically normalizing the fields, only one mode turns out to be massive. For the final action, we get:
\begin{align}\label{finact}
\nonumber  S = \int d^4 x  & \left[ - \frac{1}{4} \mathfrak{a}^{\mu \nu} \mathcal{E}^{\alpha \beta}_{\mu \nu} \mathfrak{a}_{\alpha \beta}
- \frac{m_{\textmd{eff}}^2}{4} \mathfrak{a}^{\mu \nu} \mathcal{F}^{\alpha \beta}_{\mu \nu} \mathfrak{a}_{\alpha \beta} - \right. \\
  {} & \left. - \frac{1}{4} \mathfrak{b}^{\mu \nu} \mathcal{E}^{\alpha \beta}_{\mu \nu} \mathfrak{b}_{\alpha \beta}
+ \frac{t_a}{ 2 M_h } \mathfrak{a}_{\mu \nu} T^{\mu \nu}
+ \frac{t_b}{ 2 M_h } \mathfrak{b}_{\mu \nu} T^{\mu \nu} \right].
\end{align}
\be
\mu^2 \equiv   \frac{\kappa}{\kappa - \beta} \cdot \frac{\gamma + \kappa  (\alpha +\kappa ) }{\gamma -\kappa  \left(\alpha -\beta +
\frac{\alpha ^2}{\kappa -\beta } \right)}.
\ee
Here we defined the effective mass as $m_{\textmd{eff}}^2 =  \mu^2 m^2$. The parameters $t_{a/b}$ are functions of $\alpha,\beta,\gamma$ and $\kappa$. Considering the constraints in (\ref{cons}), it's trivial to show that $m_{\textmd{eff}}^2 > 0$. The EoM's corresponding to (\ref{finact}) are:
\be\label{lin_EoM}
\mathcal{E}^{\alpha \beta}_{\mu \nu} \mathfrak{a}_{\alpha \beta} + m_{\mathrm{eff}}^2 \mathcal{F}^{\alpha \beta}_{\mu \nu} \mathfrak{a}_{\alpha \beta} =
\frac{t_a}{M_h} T_{\mu \nu},  \5\5\5
\mathcal{E}^{\alpha \beta}_{\mu \nu} \mathfrak{b}_{\alpha \beta} = \frac{t_b}{ M_h} T_{\mu \nu}.
\ee
Assuming that the source $T^{\mu \nu}$ is conserved and imposing the linear de Donder gauge for the $\mathfrak{b_{\mu \nu}}$, (\ref{lin_EoM}) reduce to:
\begin{align}
  \left( \Box - m_{\mathrm{eff}}^2 \right) \mathfrak{a}_{\mu \nu} = &  - \frac{2 t_a}{ M_h}
\left( T_{\mu \nu} - \frac{1}{3} T \eta_{\mu \nu} + \frac{1}{3 m_{\mathrm{eff}}^2} \partial_\mu \partial_\nu T \right), \\
  \Box \mathfrak{b}_{\mu \nu} = &   - \frac{2 t_b}{ M_h} \left( T_{\mu \nu} - \frac12 T \eta_{\mu \nu} \right).
\end{align}
For the exchange amplitude between two sources (Fig. \ref{feyn}) we get the sum:
\begin{align}
\nonumber  A^{tot} = & \textcolor[rgb]{1.00,0.00,0.00}{- \frac{2 | t_a |}{ M_h} \int d^4 x
\tilde{T}^{\mu \nu} \frac{1}{ \Box - m_{\mathrm{eff}}^2 }
\left( T_{\mu \nu} - \frac{1}{3} T \eta_{\mu \nu} + \frac{1}{3 m_{\mathrm{eff}}^2} \partial_\mu \partial_\nu T \right) } \\
  {} & \textcolor[rgb]{0.00,0.00,0.55}{- \frac{2 | t_b |}{ M_h} \int d^4 x
\tilde{T}^{\mu \nu} \frac{1}{ \Box }
\left( T_{\mu \nu} - \frac{1}{2} T \eta_{\mu \nu}  \right)}.
\end{align}

%%%%%%%%%%%%%%%%%%%%%%%%%%%%%%%%%%%%%%%%%%%%%%%%%%%%%%%%%%%%%%%%%%%%%%%%%%
%%%%%%%%%%%%%%%%%%%%%%%%%%%%%%%%%%%%%%%%%%%%%%%%%%%%%%%%%%%%%%%%%%%%%%%%%%
%%%%%%%%%%%%%%%%%%%%%%%%%%%%%%%%%%%%%%%%%%%%%%%%%%%%%%%%%%%%%%%%%%%%%%%%%%

\subsection{Decoupling Limit}

%%%%%%%%%%%%%%%%%%%%%%%%%%%%%%%%%%%%%%%%%%%%%%%%%%%%%%%%%%%%%%%%%%%%%%%%%%
%%%%%%%%%%%%%%%%%%%%%%%%%%%%%%%%%%%%%%%%%%%%%%%%%%%%%%%%%%%%%%%%%%%%%%%%%%
%%%%%%%%%%%%%%%%%%%%%%%%%%%%%%%%%%%%%%%%%%%%%%%%%%%%%%%%%%%%%%%%%%%%%%%%%%

The full $\Lambda = (m^2 M_h)^{1/3}$ decoupling limit (dl) of (\ref{EbG}), with parameters $\alpha=\beta=0$ and $\gamma=1$, was studied in \cite{Fasiello:2013woa}. Before I move to general case I'll try to briefly review their results. Neglecting the vector modes the dl bi-Gravity action is:
\begin{align}
\nonumber   S_{bG} = & \int d^4 x \left[ - \frac{1}{4} h^{\mu \nu} \mathcal{E}_{\mu \nu}^{\alpha \beta} h_{\alpha \beta}  - \frac{1}{4} f^{\mu \nu} \mathcal{E}_{\mu \nu}^{\alpha \beta} f_{\alpha \beta} + \right. \\
{} & + \frac{1}{8} h^{\mu \nu} \left( 2 X^{(1)}_{\mu \nu} - \frac{1}{\Lambda^3} (2 + 3 \alpha_3) X^{(2)}_{\mu \nu} + \frac{1}{\Lambda^6} ( \alpha_3 + 4 \alpha_4 ) X^{(3)}_{\mu \nu} \right)  \\
\nonumber  {} & + \left. \frac{\kappa}{8} f^{\mu \nu} \left( 2 Y^{(1)}_{\mu \nu} + \frac{1}{\Lambda^3} (4 + 3 \alpha_3) Y^{(2)}_{\mu \nu} + \frac{1}{\Lambda^6} ( 2 + 3 \alpha_3 + 4 \alpha_4 ) Y^{(3)}_{\mu \nu} \right) \right].
\end{align}
\be
X^{ \mu \nu}_{(n)} \equiv
-\frac{1}{(3-n)!} \varepsilon^{\mu \mu_1 \ldots \mu_n \sigma_1 \ldots \sigma_{3-n}}
{\varepsilon^{\nu \nu_1 \ldots \nu_n}}_{ \sigma_1 \ldots \sigma_{3-n}}
\partial_{\mu_1} \partial_{\nu_1} \pi \cdots \partial_{\mu_n} \partial_{\nu_n} \pi.
\ee
Here $Y^{(i)}_{\mu \nu}$ are defined in a fashion similar to $X^{(i)}_{\mu \nu}$ with $\pi \rightarrow \rho$, where
$\rho(x)$ is the Galileon field from the point of view of the $\tilde{g}$ metric and is related non-locally to $\pi(x)$:
\be
\rho (x) = - \pi (x) + \frac{1}{2 \Lambda^3} \left( \partial_\mu \pi (x) \right)^2 - \frac{1}{2 \Lambda^6} \partial^\mu \pi (x) \partial^\nu \pi (x) \partial_\mu \partial_\nu \pi (x) + \cdots
\ee
Although this definition involves higher derivatives, because of its nonlocal structure, the Ostrogradski ghost is absent. Making the field redefinitions,
\begin{align}\label{field_redef}
h_{\mu \nu} = &  \hat{h}_{\mu \nu} + \eta_{\mu \nu} \pi + \frac{1}{2 \Lambda^3} \left( 2 + 3 \alpha_3 \right) \partial_\mu \pi \partial_\nu \pi,   \\
f_{\mu \nu} = & \hat{f}_{\mu \nu} + \kappa \eta_{\mu \nu} \rho - \frac{\kappa}{2 \Lambda^3} \left( 4 + 3 \alpha_3 \right) \partial_\mu \rho \partial_\nu \rho,
\end{align}
we partially unmix the tensor modes from the scalars. After unmixing, the bi-Gravity Action becomes:
\begin{align}
\nonumber S_{bG}  = &  \int d^4 x \left[ - \frac{1}{4} \hat{h}^{\mu \nu} \mathcal{E}_{\mu \nu}^{\alpha \beta} \hat{h}_{\alpha \beta}
- \frac{1}{4} \hat{f}^{\mu \nu} \mathcal{E}_{\mu \nu}^{\alpha \beta} \hat{f}_{\alpha \beta} -  \right. \\
	   {} &  - \frac{1}{8 \Lambda^6} \left( \alpha_3 + 4 \alpha_4 \right)  \hat{h}^{\mu \nu} X_{\mu \nu}^{(3)} + \frac{\kappa}{8 \Lambda^6} \left( 2 + 3 \alpha_3 + 4 \alpha_4 \right)  \hat{f}^{\mu \nu} Y_{\mu \nu}^{(3)} + \\
\nonumber  {} & + \sum_{n=2}^{5} \frac{c_n}{\Lambda^{3(n-2)}} \mathcal{L}^{(n)}_{\text{gal}} \left[ \pi \right] +
 \sum_{n=2}^{5} \left. \frac{\tilde{c}_n}{\Lambda^{3(n-2)}} \mathcal{L}^{(n)}_{\text{gal}} \left[ \rho \right] \right].
\end{align}
\indent We are now going to study the $\Lambda = (m^2 M_h)^{1/3}$ decoupling limit of the theory defined by (\ref{EbG}). We start by introducing the St\"uckelberg fields which recover the full diffeomorphism invariance $diff \otimes \widetilde{diff}$. According to the St\"uckelberg trick we make the following substitutions (choose to St\"uckelberguise $\tilde{g}_{\mu \nu}$):
\be
\tilde{g}_{\alpha \beta} (x) \longrightarrow  \tilde{g}_{a b} (\Phi)
\frac{\partial \Phi^a}{\partial x^\alpha} \frac{\partial \Phi^b}{\partial x^\beta}.
\ee
Now the theory is invariant under the two copies of the diffeomorphisms. Note that we attributed the Greek index to $diff$ and the Latin index to $\widetilde{diff}$. The vectors from the $diff$ and $\widetilde{diff}$ sectors transform respectively as:
\be
A_\mu \longrightarrow A'_\mu = \frac{\partial x^\alpha}{\partial y^\mu} A_\alpha,   \5\5\5
\Psi_a \longrightarrow \Psi'_a = \frac{\partial \Phi^i}{\partial Y^a} \Psi_i.
\ee
After the St\"uckelberg substitution, the Lagrangian density for the new kinetic terms takes the form:
\be\label{kinLag}
\Delta \mathcal{L}  = (\alpha - \beta ) \frac{M_h M_f}{2} \sqrt{-g} g^{\mu \nu} \frac{\partial \Phi^a}{\partial x^\mu} \frac{\partial \Phi^b}{\partial x^\nu} \tilde{R}_{a b} +
\beta  \frac{M_h M_f}{2} \sqrt{-\tilde{g}} \left| \frac{\partial \Phi}{\partial x} \right| \tilde{g}^{a b}
 \frac{\partial x^\mu}{\partial \Phi^a}  \frac{\partial x^\nu}{\partial \Phi^b}  R_{\mu \nu}.\footnote{This is the only way to recover two copies of diffeomorphism invariance without introducing new fields and without explicitly breaking the Poincar\'{e} symmetry.}
\ee
Next we define the canonically normalized fields \cite{Gabadadze:2013ria}:
\begin{align}
\nonumber  g_{\mu \nu} = & \eta_{\mu \nu} + \frac{1}{M_h} h_{\mu \nu}, \\
  \tilde{g}_{a b} = & \eta_{a b} + \frac{1}{M_f} f_{a b}, \\
\nonumber  \frac{\partial \Phi^a}{\partial x^\mu} = & \partial_\mu \left( x^a + \frac{m}{\Lambda^3} B^a +\frac{1}{\Lambda^3} \partial^a \pi \right)
\end{align}
and take the following limit:
\be\label{dec_lim_def}
M_h \rightarrow +\infty, \5\5 M_f \rightarrow +\infty, \5\5 m \rightarrow 0, \5\5 \Lambda \rightarrow \text{const}, \5\5 \kappa \rightarrow \text{const}.
\ee
The parameters $\alpha,\beta,\gamma$ are kept constant in the decoupling limit. As we will see these two terms contain parts that become strongly coupled in the (\ref{dec_lim_def}) limit and host the BD ghost as well. The complete decoupling limit for these terms:
\begin{align}\label{first}
\nonumber M_g M_f & \sqrt{-g}  g^{\mu \nu} \frac{\partial \Phi^a}{\partial x^\mu}
  \frac{\partial \Phi^b}{\partial x^\nu} \tilde{R}_{a b} =
\textcolor[rgb]{1.00,0.00,0.00}{M_g} \eta^{\mu \nu} \tilde{R}_{a b}^{(10)} \phi_0{}_\mu^a \phi_0{}_\nu^b +
\frac{\Lambda^3}{\textcolor[rgb]{1.00,0.00,0.00}{m}} \eta^{\mu \nu} \left( 2 \tilde{R}_{a b}^{(10)} \phi_0{}_\mu^a \phi_1{}_\nu^b \right. + \\
\nonumber {} & + \left. \tilde{R}_{a b}^{(11)} \phi_0{}_\mu^a \phi_0{}_\nu^b \right)
 + \frac12 h \eta^{\mu \nu} \tilde{R}_{a b}^{(10)} \phi_0{}_\mu^a \phi_0{}_\nu^b -
h^{\mu \nu} \tilde{R}_{a b}^{(10)} \phi_0{}_\mu^a \phi_0{}_\nu^b +
 \kappa \eta^{\mu \nu} \tilde{R}_{a b}^{(20)} \phi_0{}_\mu^a \phi_0{}_\nu^b + \\
{} & + \Lambda^3 \eta^{\mu \nu} \left( \tilde{R}_{a b}^{(10)} \phi_1{}_\mu^a \phi_1{}_\nu^b + 2 \tilde{R}_{a b}^{(11)} \phi_1{}_\mu^a \phi_0{}_\nu^b + \tilde{R}_{a b}^{(12)} \phi_0{}_\mu^a \phi_0{}_\nu^b \right),
\end{align}
\begin{align}\label{second}
\nonumber M_g M_f & \sqrt{-\tilde{g}}  \left| \frac{\partial \Phi}{\partial x} \right| \tilde{g}^{a b}
\frac{\partial x^\mu}{\partial \Phi^a} \frac{\partial x^\nu}{\partial \Phi^b}  R_{\mu \nu} =
\textcolor[rgb]{1.00,0.00,0.00}{M_f} \eta^{a b} R_{\mu \nu}^{(1)} \psi_0{}_{a b}^{\mu \nu} +
 \frac{\Lambda^3}{\kappa \textcolor[rgb]{1.00,0.00,0.00}{m}} \eta^{a b} R_{\mu \nu}^{(1)} \psi_1{}_{a b}^{\mu \nu} +  \\
  {} & + \frac12 f \eta^{a b} R_{\mu \nu}^{(1)} \psi_0{}_{a b}^{\mu \nu} -
f^{a b} R_{\mu \nu}^{(1)} \psi_0{}_{a b}^{\mu \nu} +
 \frac{1}{\kappa} \eta^{a b} R_{\mu \nu}^{(2)} \psi_0{}_{a b}^{\mu \nu} +
\frac{\Lambda^3}{\kappa} \eta^{a b} R_{\mu \nu}^{(1)} \psi_2{}_{a b}^{\mu \nu}.
\end{align}

In addition to the $f_{a b}$, the Ricci tensor $\tilde{R}_{a b} = \tilde{R}_{a b} \left( \Phi \right)$ also depends on the vector $(B^a)$ and the scalar $(\pi)$ modes. In fact wrt the scalar mode it's an infinite series. One should also keep in mind that $f_{a b} \left( \Phi \right) = f_{a b} \left( x \right) + \frac{1}{\Lambda^3} \partial^c \pi \partial_c  f_{a b} + \frac{1}{2\Lambda^6} \partial^c \pi \partial^d \pi \partial_c \partial_d  f_{a b} + \cdots + \text{vector modes} $. The first upper index on the rhs of (\ref{first}) and (\ref{second}) (in parenthesis) corresponds to the order of spin 2 field, while the second upper index indicates the order of the spin 1 field $(B^a)$. Lower indices indicate the order of spin 1 field. Other notations:
\begin{align}\label{notations}
  \frac{\partial \Phi^a}{\partial x^\mu} \equiv  & \phi_0{}_\mu^a + m \cdot \phi_1{}_\mu^a, \\
  \frac{\partial x^\mu}{\partial \Phi^a} \equiv  & \tilde{\phi}_0{}_a^\mu + m \cdot \tilde{\phi}_1{}_a^\mu + m^2 \cdot \tilde{\phi}_2{}_a^\mu + \mathcal{O} (m^3), \\
  \left| \frac{\partial \Phi}{\partial x} \right| \equiv & \phi_0 + m \cdot \phi_1 + m^2 \cdot \phi_2, \\
  \psi_0{}_{a b}^{\mu \nu} \equiv & \phi_0 \tilde{\phi}_0{}_a^\mu \tilde{\phi}_0{}_b^\nu, \\
  \psi_1{}_{a b}^{\mu \nu} \equiv  & \phi_1 \tilde{\phi}_0{}_a^\mu \tilde{\phi}_0{}_b^\nu +  \phi_0 \tilde{\phi}_0{}_a^\mu \tilde{\phi}_1{}_b^\nu
  +  \phi_0 \tilde{\phi}_1{}_a^\mu \tilde{\phi}_0{}_b^\nu, \\
  \psi_2{}_{a b}^{\mu \nu} \equiv & \phi_0  \tilde{\phi}_1{}_a^\mu \tilde{\phi}_1{}_b^\nu +
  \phi_0 \tilde{\phi}_0{}_a^\mu \tilde{\phi}_2{}_b^\nu  + \phi_0 \tilde{\phi}_2{}_a^\mu \tilde{\phi}_0{}_b^\nu  +
  \phi_1 \tilde{\phi}_0{}_a^\mu \tilde{\phi}_1{}_b^\nu + \phi_1 \tilde{\phi}_1{}_a^\mu \tilde{\phi}_0{}_b^\nu +
\phi_2 \tilde{\phi}_0{}_a^\mu \tilde{\phi}_0{}_b^\nu.
\end{align}

\indent The highlighted terms in (\ref{first}) and (\ref{second}) are not total derivatives and become strongly coupled in the limit (\ref{dec_lim_def}). This is the first problem encountered within this model. From now on we will forget the existence of strongly coupled terms, since they are irrelevant for the rest of our analysis. \\

%%%%%%%%%%%%%%%%%%%%%%%%%%%%%%%%%%%%%%%%%%%%%%%%%%%%%%%%%%%%%%%%%%%%%%%%%%
%%%%%%%%%%%%%%%%%%%%%%%%%%%%%%%%%%%%%%%%%%%%%%%%%%%%%%%%%%%%%%%%%%%%%%%%%%
%%%%%%%%%%%%%%%%%%%%%%%%%%%%%%%%%%%%%%%%%%%%%%%%%%%%%%%%%%%%%%%%%%%%%%%%%%

\subsection{Analysis in the Decoupling Limit}

%%%%%%%%%%%%%%%%%%%%%%%%%%%%%%%%%%%%%%%%%%%%%%%%%%%%%%%%%%%%%%%%%%%%%%%%%%
%%%%%%%%%%%%%%%%%%%%%%%%%%%%%%%%%%%%%%%%%%%%%%%%%%%%%%%%%%%%%%%%%%%%%%%%%%
%%%%%%%%%%%%%%%%%%%%%%%%%%%%%%%%%%%%%%%%%%%%%%%%%%%%%%%%%%%%%%%%%%%%%%%%%%

\indent Let's try to prove the existence of the ghost. We first unmix the scalar from the spin-2 modes at the quadratic level. Assuming that at the beginning only $h_{\mu \nu}$ is coupled to the source, the linear Lagrangian after the unmixing becomes:
\begin{align}
\nonumber  \mathcal{L} = & - \frac{1}{4}
\left( \begin{array}{c} h^{\mu \nu} \\ f^{\mu \nu} \end{array} \right)^T
\left( \begin{array}{cc} 1 - \beta/ \kappa & \alpha \\
\alpha & \gamma - \alpha \kappa + \beta \kappa \end{array} \right) \mathcal{E}^{\alpha \beta}_{\mu \nu}
\left( \begin{array}{c} h_{\alpha \beta} \\ f_{\alpha \beta} \end{array} \right) - \\
  {} & - \frac{3}{4} \mu^2 \left( \partial \pi \right)^2 + \frac{1}{2 M_h} h^{\mu \nu} T_{\mu \nu} + \frac{a_1}{2 M_h} \pi T.
\end{align}
Let's take a static and spherically symmetric source with the following ansatz:
\be
T_{\mu \nu} = s^4 \theta \left( r_* - r \right) \eta_{\mu 0} \eta_{\nu 0}.
\ee
Classical solutions of this theory:
\be
h_{0 0 }^{\text{in}} (r) = \frac{b_2 s^4}{2 M_h} \left( \frac{r^2}{3} - r_*^2 \right),    \5\5\5
h_{0 0 }^{\text{out}} (r) = - \frac{b_2 s^4 r_*^3}{3 M_h r},
\ee
\be
f_{0 0 }^{\text{in}} (r) = \frac{b_1 s^4}{2 M_h} \left( \frac{r^2}{3} - r_*^2 \right),    \5\5\5
f_{0 0 }^{\text{out}} (r) = - \frac{b_1 s^4 r_*^4}{3 M_h r},
\ee
\be
\pi^{\text{in}} (r) = \frac{a_1 s^4}{6 \mu^2 M_h} \left( \frac{r^2}{3} - r_*^2 \right),    \5\5\5
\pi^{\text{out}} (r) = - \frac{ a_1 s^4 r_*^3}{9 \mu^2 M_h r}.
\ee
for which we introduce the following notations:
\begin{align}
  a_1 =  \frac{\gamma +\kappa  \beta }{\gamma +\kappa \left( \alpha +\kappa \right)} \mu^2, & \5\5\5
  a_2 =  \frac{\beta - \alpha - \kappa }{\gamma +\kappa \left( \alpha +\kappa \right)} \mu^2, \\
  b_1 = - \frac{\alpha}{\gamma +\kappa \left( \alpha +\kappa \right)} \mu^2, & \5\5\5
  b_2 =  \frac{\gamma +\kappa \left( \beta - \alpha  \right)}{\gamma +\kappa \left( \alpha +\kappa \right)} \mu^2.
\end{align}
Let's assume that we are inside the source near the wall ($r<r_*$ and $r\approx r_*$) and study the fluctuation of the scalar field, while freezing the tensors to their classical values:
\be
\pi = \pi_c + \bar{\pi}.
\ee
Corrections to the kinetic term (those that might excite the ghost) coming from the cubic level ($\Pi_{\mu \nu} = \partial_\mu \partial_\nu \pi$):
\be
\mathcal{L}^{(3)} = \frac{1}{2 \Lambda^3} \left[ \left( \alpha - \beta \right) a_2 + \frac{\beta a_1}{\kappa} \right] \times
\left[  - \frac12 h_c \left[ \bar{\Pi} \right]^2 + 2 h^{\mu \nu}_c \bar{\Pi}_{\mu \nu} \left[ \bar{\Pi} \right]  -
\kappa \left(  h_c \rightarrow f_c  \right)  \right].
\ee
Note that the $3\pi$ vertex is a total derivative. After the derivation of the EoM's we find four time derivatives acting on $\bar{\pi}$ and therefore giving rise to the instability with the mass:
\be
M_{\text{ghost}}^2 = - \frac{3 \mu^2}{a_1} \left[ \left( \alpha - \beta \right) a_2 + \frac{\beta a_1}{\kappa} \right]^{-1} \frac{1}{r^2_*} \frac{M_h}{s} \left(\frac{\Lambda}{s}\right)^3.
\ee
In order for this solution to be valid we need to set three conditions: \\
\ding{182} Condition for the next order corrections to be small:
\be
\frac{s^4}{M_h \Lambda^3} \ll 1.
\ee
\ding{183} Condition for the mass of the ghost to be small compared to the cutoff scale:
\be
\frac{1}{r_*^2} \cdot \frac{M_h \Lambda^3}{s^4} \ll \Lambda^2.
\ee
\ding{184} Condition for the absence of the black hole:
\be
\frac{s^4 r_*^2}{ M_h } \ll M_h.
\ee
It is possible to satisfy all these conditions. For instance:
\be
M_h = 10^{19} GeV, \5\5 s = 10^{-2} GeV,  \5\5  r_* = 10^{15} GeV^{-1},  \5\5  \Lambda = 10^{-7} GeV,
\ee
or in more familiar units, the mass of the source $=s^4 r_*^3 = 10^{10} Kg$; radius of the source $r_* = 10 cm$ and Schwarzschild radius for the source $r_g = 10^{-15} cm$. \\
\indent We can avoid this ghost by setting
\be\label{ghost_freedom1}
\left( \alpha - \beta \right) a_2 + \frac{\beta a_1}{\kappa} = 0,
\ee
but in this case it will re-emerge at the quartic level. To make this clear let's forget about the source and consider the following exact solution to the classical equations of motion:
\be
h^{\mu \nu}_c =0, \5\5 f^{\mu \nu}_c =0, \5\5 \pi_c = \frac12 \left( a \cdot x \right)^2,
\ee
where $a_\mu = \left( a_0, -\mathbf{a}_0 \right)$ is some light-like vector. Since there are no $3 \pi $ interactions at the cubic level, the scalar kinetic term will start receiving corrections from the quartic order:
\begin{align}
\nonumber   \mathcal{L}^{(4)}_\pi = & \frac{1}{2 \Lambda^6} a_2  \left( \alpha - \beta \right) \left(  \pi_c \Pi_c^{\mu \nu} \left[ \Pi \right] \Pi_{\mu \nu} - 6 \pi_c \Pi_c^{\mu \nu}\Pi^2_{\mu \nu} - \frac{9}{2} \partial^\mu \pi_c \partial^\nu \pi_c \Pi^2_{\mu \nu} \right) \mu^2 -   \\
\nonumber  {} & - \frac{1}{2 \Lambda^6} a_1 \frac{\beta }{\kappa}  \left( \pi_c \Pi_c^{\mu \nu} \left[ \Pi \right] \Pi_{\mu \nu} - 6 \pi_c \Pi_c^{\mu \nu}\Pi^2_{\mu \nu} - \frac{3}{2} \partial^\mu \pi_c \partial^\nu \pi_c \Pi^2_{\mu \nu} \right) \mu^2.
\end{align}
The mass of the ghost in this case is:
\be
M_{\text{ghost}}^2 = - \frac{3  }{2 } \left[ 14 \left( \alpha - \beta \right) a_2 - 8 \frac{\beta a_1}{\kappa} \right]^{-1} \frac{\Lambda^6}{a_0^2 \left( a \cdot x \right)^2}.
\ee
To justify our approximation we set: \\
\ding{172} Condition for the possible next order corrections to be small:
\be
\frac{a_0^2}{ \Lambda^3} \ll 1.
\ee
\ding{173} Condition for the mass to be small compared to the cutoff scale:
\be
\frac{\Lambda^6}{a_0^2 \left( a \cdot x \right)^2} \ll \Lambda^2.
\ee
By properly choosing $a_0$ and $x$ it is possible to fulfill both of these conditions.  \\
We can get rid of this ghost by setting:
\be\label{ghost_freedom2}
14 \left( \alpha - \beta \right) a_2 - 8 \frac{\beta a_1}{\kappa} = 0,
\ee
but this condition is not consistent with (\ref{ghost_freedom1}), since in order for both, (\ref{ghost_freedom1}) and (\ref{ghost_freedom2}) to hold we need
\be\label{consistency}
\alpha = \beta
\ee
and this condition goes against our assumptions. At this point one might still think that $\alpha = \beta \neq 0$ might be a solution. Note that the Lagrangian (\ref{EbG}) without the matter part and dRGT potential is invariant under $ g \leftrightarrow \tilde{g} $ (up to a redefinition of constants). As a consequence of this fact if we go back and St\"uckelberguise the field $g$ instead of $\tilde{g}$, replace $ g \rightarrow \tilde{g} $ in the matter part and run the same analysis the consistency condition (\ref{consistency}) will be replaced with $\beta=0$. \\

%%%%%%%%%%%%%%%%%%%%%%%%%%%%%%%%%%%%%%%%%%%%%%%%%%%%%%%%%%%%%%%%%%%%%
%%%%%%%%%%%%%%%%%%%%%%%%%%%%%%%%%%%%%%%%%%%%%%%%%%%%%%%%%%%%%%%%%%%%%
%%%%%%%%%%%%%%%%%%%%%%%%%%%%%%%%%%%%%%%%%%%%%%%%%%%%%%%%%%%%%%%%%%%%%

\section{Conclusions and Discussion}

We proposed necessary criteria for derivative mixings of spin-2 fields to be ghost free and studied few examples (\ref{EbG}). We proved that for any choice of parameters ($\alpha$,
$\beta$ and $\gamma$) the nonlinear model hosts a ghost. The Action (\ref{EbG}) is not the most general, in fact as we mentioned there exist infinitely many terms that satisfy criteria \ref{crit1},\ref{crit2} and \ref{crit3}. These infinitely many terms bring infinitely many parameters into the action and this makes it hard to prove the existence of ghost in the general case.

%%%%%%%%%%%%%%%%%%%%%%%%%%%%%%%%%%%%%%%%%%%%%%%%%%%%%%%%%%%%%%%%%%%%%
%%%%%%%%%%%%%%%%%%%%%%%%%%%%%%%%%%%%%%%%%%%%%%%%%%%%%%%%%%%%%%%%%%%%%
%%%%%%%%%%%%%%%%%%%%%%%%%%%%%%%%%%%%%%%%%%%%%%%%%%%%%%%%%%%%%%%%%%%%%

\acknowledgments

I'm indebted to Prof. Gregory Gabadadze for proposing this project and guiding me through it. I would like to thank Anna-Maria Taki and Siqing Yu for the feedback on a draft version of this paper. The project was supported by the NYU MacCracken scholarship and by the NSF grant PHY-1316452.

%%%%%%%%%%%%%%%%%%%%%%%%%%%%%%%%%%%%%%%%%%%%%%%%%%%%%%%%%%%%%%%%%%%%%
%%%%%%%%%%%%%%%%%%%%%%%%%%%%%%%%%%%%%%%%%%%%%%%%%%%%%%%%%%%%%%%%%%%%%
%%%%%%%%%%%%%%%%%%%%%%%%%%%%%%%%%%%%%%%%%%%%%%%%%%%%%%%%%%%%%%%%%%%%%

\appendix
\section{Appendix: Equations of Motion}

\noindent The equations of motion corresponding to (\ref{EbG})
\begin{align}
\nonumber {} & \frac{1}{\sqrt{-g}} \frac{\delta S_{EbG} }{\delta g^{\mu \nu}} =  R_{\mu \nu} - \frac12 g_{\mu \nu} R + \frac{\alpha - \beta}{ \kappa } \left[ \tilde{R}_{\mu \nu} - \frac12 g_{\mu \nu} g^{\alpha \beta} \tilde{R}_{\alpha \beta} \right] + \\
 {} & + \frac{\beta}{ 2\kappa} \left[  \nabla_\alpha \nabla_\beta \left( a \tilde{g}^{\alpha \beta} g_{\mu \nu} \right) +
 g^{\alpha \beta} \nabla_\alpha \nabla_\beta \left( a \tilde{g}^{\lambda \rho} g_{\mu \rho} g_{\nu \lambda} \right) -
  \nabla_\alpha \nabla_\mu \left( a \tilde{g}^{\alpha \beta}  g_{\nu \beta} \right) -  \right. \\
\nonumber {} &  \left. - \nabla_\alpha \nabla_\nu \left( a \tilde{g}^{\alpha \beta}  g_{\mu \beta} \right) \right]  + \frac{m^2 }{2} U_{\mu \nu} + \frac{1}{M_h^2} T_{\mu \nu} = 0,
\end{align}
\begin{align}
\nonumber {} & \frac{1}{\sqrt{-\tilde{g}}}  \frac{\delta S_{EbG} }{\delta \tilde{g}^{\mu \nu}}  = \tilde{R}_{\mu \nu} - \frac12 \tilde{g}_{\mu \nu} \tilde{R} +
 \frac{\beta ~ \kappa}{ \gamma } \left[ R_{\mu \nu} - \frac12 \tilde{g}_{\mu \nu} \tilde{g}^{\alpha \beta} R_{\alpha \beta} \right] + \\
 {} & + \frac{\kappa ~ (\alpha - \beta)}{ 2 \gamma} \left[  \tilde{\nabla}_\alpha \tilde{\nabla}_\beta \left( \frac{1}{a} g^{\alpha \beta} \tilde{g}_{\mu \nu} \right) +
 \tilde{g}^{\alpha \beta} \tilde{\nabla}_\alpha \tilde{\nabla}_\beta \left( \frac{1}{a} g^{\lambda \rho} \tilde{g}_{\mu \rho} \tilde{g}_{\nu \lambda} \right) - \right. \\
\nonumber {} & \left. - \tilde{\nabla}_\alpha \tilde{\nabla}_\mu \left( \frac{1}{a} g^{\alpha \beta}  \tilde{g}_{\nu \beta} \right) -
\tilde{\nabla}_\alpha \tilde{\nabla}_\nu \left( \frac{1}{a} g^{\alpha \beta}  \tilde{g}_{\mu \beta} \right) \right] +  \frac{m^2 \kappa^2}{2 \gamma} \tilde{U}_{\mu \nu} = 0.
\end{align}
Here $a^2 \equiv \tilde{g}/g$, $\nabla (\tilde{\nabla})$ stands for the covariant derivative wrt $g (\tilde{g})$, $T_{\mu \nu}$ is the Energy momentum tensor for the matter fields and $U (\tilde{U})$ are defined as:
\be
\frac{1}{\sqrt{-g}} \frac{\delta S_{dRGT} }{\delta g^{\mu \nu}}  \equiv \frac{m^2 M_{h}^2}{4} U_{\mu \nu},
\ee
\be
\frac{1}{\sqrt{-\tilde{g}}} \frac{\delta S_{dRGT} }{\delta \tilde{g}^{\mu \nu}} \equiv \frac{m^2 M_{h}^2}{4} \tilde{U}_{\mu \nu}.
\ee

%%%%%%%%%%%%%%%%%%%%%%%%%%%%%%%%%%%%%%%%%%%%%%%%%%%%%%%%%%%%%%%%%%%%%
%%%%%%%%%%%%%%%%%%%%%%%%%%%%%%%%%%%%%%%%%%%%%%%%%%%%%%%%%%%%%%%%%%%%%
%%%%%%%%%%%%%%%%%%%%%%%%%%%%%%%%%%%%%%%%%%%%%%%%%%%%%%%%%%%%%%%%%%%%%

%%%%%%%%%%%%%%%%%%%%%%%%%%%%%%%%%%%%%%%%%%%%%%%%%%%%%%%%%%%%%%%%%%%%%
%%%%%%%%%%%%%%%%%%%%%%%%%%%%%%%%%%%%%%%%%%%%%%%%%%%%%%%%%%%%%%%%%%%%%
%%%%%%%%%%%%%%%%%%%%%%%%%%%%%%%%%%%%%%%%%%%%%%%%%%%%%%%%%%%%%%%%%%%%%

\end{document}